\documentclass[lettersize,journal]{IEEEtran}
\usepackage{amsmath,amsfonts}
\usepackage{algorithmic}
\usepackage{array}
\usepackage[caption=false,font=normalsize,labelfont=sf,textfont=sf]{subfig}
\usepackage{textcomp}
\usepackage{stfloats}
\usepackage{url}
\usepackage{verbatim}
\usepackage{graphicx}
\usepackage{authblk}
\usepackage{orcidlink}

\def\BibTeX{{\rm B\kern-.05em{\sc i\kern-.025em b}\kern-.08em
    T\kern-.1667em\lower.7ex\hbox{E}\kern-.125emX}}
\usepackage{balance}
\begin{document}
\title{LANTERN: Characterization technology for low threshold cryogenic detectors}
\author[1]{G. Del Castello}
\affil[1]{INFN Sezione di Roma, P.le A. Moro 2, 00185 Roma, Italy}

\markboth{}%
{}

\maketitle

\begin{abstract}
The use of low-temperature detectors, such as cryogenic calorimeters, has pioneered the recent advancements in low-energy rare event searches. These detectors provide a low-noise environment essential for the direct detection of dark matter and neutrinos. Characterizing these detectors within the region of interest (ROI), typically spanning from O(10~eV) to O(1~keV), has proven to be a challenging task. Conventional radioactive sources produce signals above this range, leading to nonlinearities and saturation effects. Moreover, these detectors are usually deployed in low background environments, meaning that having a radioactive source during physics runs can spoil the measurement making the use of this type of solution unfeasible.

As a solution to these issues, we introduce LANTERN, an optical calibration system designed for highly segmented cryogenic calorimeters. LANTERN utilizes the photostatistics resulting from the absorption of monochromatic UV-Vis photons emitted by LEDs to analyze the detector response curve, without needing prior knowledge of the total energy deposited.  The system employs a fast-switching LED matrix that operates at excitation times faster than the typical response of cryogenic detectors and can currently characterize up to 64 calorimeters independently.

LANTERN induces particle-like signals on the detector in a wide energy range, from a few eV to several hundreds of keV. This allows for a complete study of the response of the detector within the ROI, including the energy calibration, cross-talk evaluation and pixel identification.

In this work, the validation of the final electronics designed for the project is shown. The first test was carried out by calibrating one of the cryogenic detectors of the BULLKID-DM experiment and checking the energy-reconstruction error of the spectral features produced by the surrounding lead casing. An error of $\approx 2\%$ has been observed in the energy reconstruction. The second validation was carried out by cross-calibrating one of the CALDER thin detectors with a commercial LED driver, and compatible results between the two setups were achieved.

LANTERN is now ready for its full deployment in the BULLKID project for R\&D purposes and detector development. The usage of LANTERN in the CRAB and NUCLEUS experiments is also currently being investigated.

\end{abstract}


\section{Introduction}
Low-temperature detectors, such as cryogenic calorimeters, are often used to study rare processes at low energies. Experiments such as CRESST \cite{cresst}, NUCLEUS \cite{NUCLEUS}, BULLKID-DM \cite{bullkid}, Cosinus \cite{cosinus}, use cryogenic calorimeters to study energy depositions in a region of interest (ROI), typically spanning from O(10 eV) to O(1 keV). Calibrating these detectors has proven to be quite a challenge due to the low-energy regime. In fact, commonly used radioactive X-ray sources deposit energies, usually of several keV, that are above the typical ROI. This means that X-ray calibration sources are affected by nonlinearities and saturation effects, resulting in an incomplete description of the detector response. Furthermore, since cryogenic calorimeters are generally operated in low-background environments, it is essential to be able to remove the calibration source without disturbing the base temperature of the cryostat. Meeting this requirement is challenging with keV-scale X-ray sources, which must be positioned inside the detector holder.

These limitations can be overcome by the use of LANTERN, an electronic and optical setup that is easily up-scaled and targeted at the low-energy calibration of cryogenic calorimeters via the use of UV-Vis photons. 

In this article, the basic principles of the optical calibration will be addressed along with a cross-calibration with X-ray peaks. Finally, a description and validation of the final LANTERN electronics will be presented by comparing the results with a simpler commercial system.

\section{Characterization of Cryogenic Calorimeters with LEDs}
	The setup of the optical calibration described below allows for several different characterizations of a cryogenic calorimeter. The main aim of the system is to provide an energy calibration at the 100~eV scale. To widen the energy range in which the calibration is valid a detector response nonlinearity evaluation can be performed, as described in Sec.~\ref{sec:nonlin}.
	\subsection{Optical Calibration}\label{sec:calib}
	
	 \begin{figure}
			\includegraphics[scale=0.35]{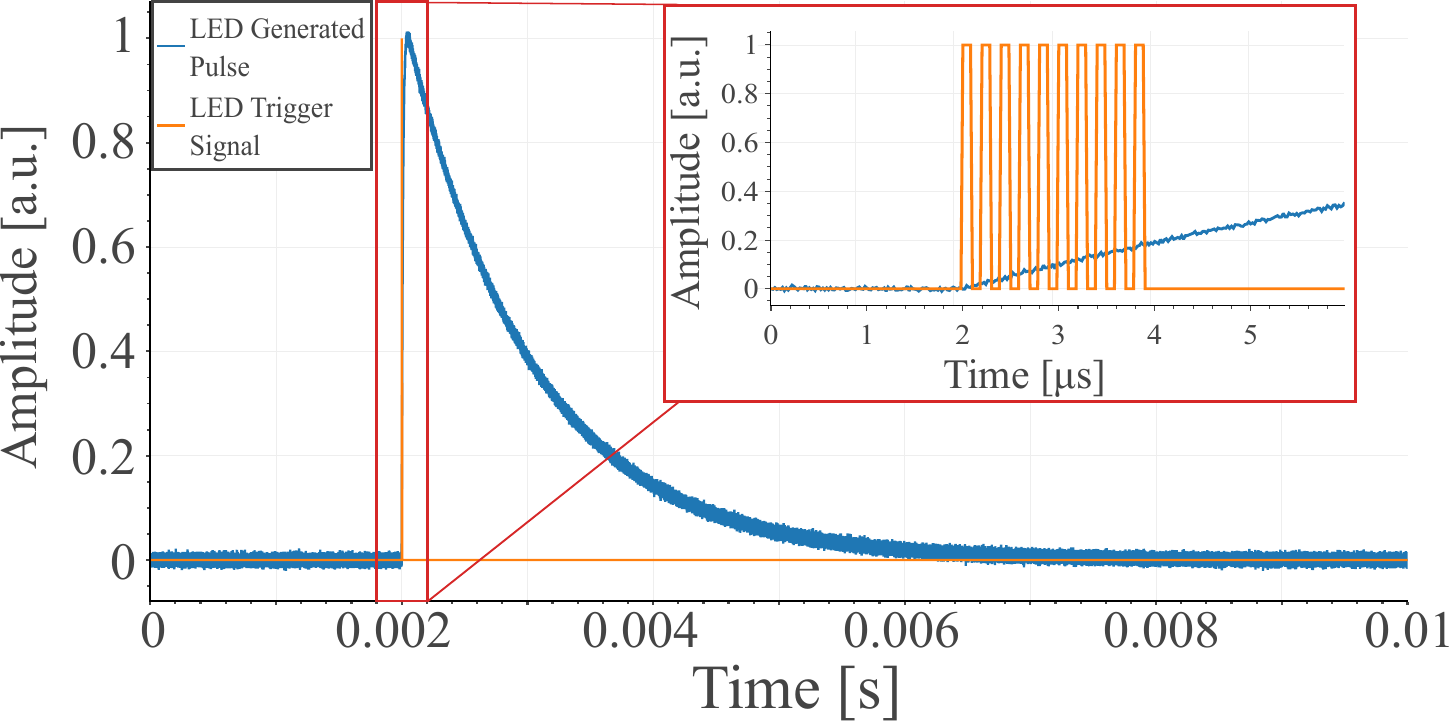}
			\caption{In blue a simulated pulse generated during an optical calibration is show along side the triggering signal sent to the LED driver plotted in orange. In the inset a zoom showing the fine structure of the trigger is presented: the 10 periods of a 5~MHz square wave is used as the trigger and each period corresponds to a single LED burst.}
			\label{fig:Simu_pulse}
		 \end{figure}	
		  
	Typical integration times of cryogenic calorimeters range from few hundred microseconds to several milliseconds, making them slower than the typical switching times achievable with a commercial LED. It is then possible to send several light pulses, referred to as bursts, to the detector such that the individual arrival times of the photons are not resolved and are instead integrated in a single response signal, as discussed in \cite{calder} and \cite{lantern1} and illustrated in Fig.~\ref{fig:Simu_pulse}. The amplitude of the LED induced pulse is then proportional to the N$\gamma$ photons absorbed by the detector, making it a Poisson distributed variable. The mean and variance of this distribution can be expressed as follows:

\begin{equation}
\mu_\gamma=r\epsilon_\gamma\left\langle N_\gamma\right\rangle \qquad \sigma^2_\gamma =\left(r\epsilon_\gamma\right)^2\left\langle N_\gamma\right\rangle = r\epsilon_\gamma\mu_\gamma
\end{equation}

where $\mu_\gamma$ and $\sigma^2_\gamma$  are respectively the mean and the variance of amplitude distribution of the generated light pulses, $r$ is the responsivity of the detector in units of detector amplitude over absorbed energy, $\epsilon_\gamma$ is the energy of the single photon (of the order of few electronvolts for UV-Vis light sources) and the $\langle\cdot\rangle$ operation is the mean. The above expressions are only valid for a detector with a negligible amount of noise, but since this calibration is aimed at detectors working near their noise level the actual measured variance of the distribution is:

\begin{equation}\label{eq:abscal_func}
\sigma^2_{LED} = \sigma^2_0 + \sigma^2_\gamma = r\epsilon_\gamma \mu_\gamma + \sigma^2_0
\end{equation}
 where $\sigma^2_0$ is the intrinsic resolution of the detector.
 
\begin{figure}
		\includegraphics[scale=0.78]{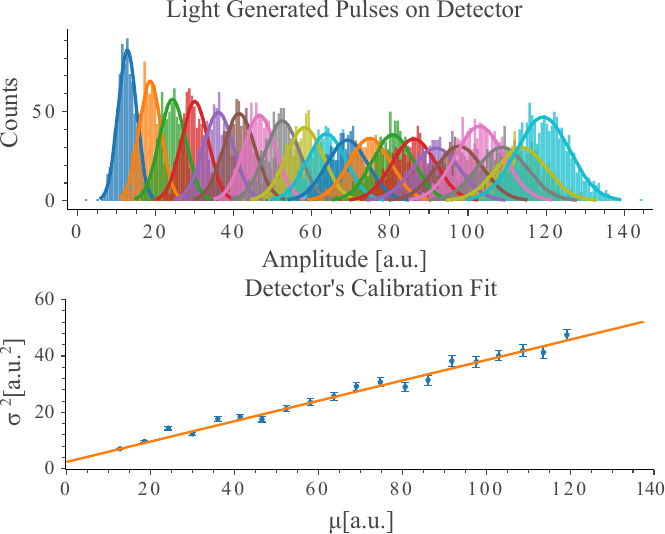}
		\caption{Example of the optical calibration procedure. In the top panel the different energy distributions of the light depositions are shown and fitted with a Gaussian. In the bottom panel, the mean and variance of the distributions are plotted and fitted with a first degree polynomial to extract the calibration parameters of Eq.~\ref{eq:abscal_func}. The figure is taken from \cite{lantern1}.}
		\label{fig:calib_example}
	\end{figure}

 As shown in Fig.~\ref{fig:calib_example} it is possible to generate multiple distributions (by varying $\langle N_\gamma\rangle$) and independently extract their mean and variance using a Gaussian fit. By fitting the behavior of $\sigma^2$ with respect to $\mu$ using Eq.~\ref{eq:abscal_func} it is possible to extract $\sigma^2_0$  and $r$ since $\epsilon_\gamma$ is known and fixed by the LED specifications. It is worth to mention that this procedure does not rely on the knowledge of $\langle N_\gamma\rangle$ meaning that it is robust with respect to unquantified light losses or yields. The only assumption made is that the cryogenic calorimeter behaves linearly with respect to $\langle N_\gamma\rangle$, which is usually verified since this calibration is meant to be performed at low energies.

		\subsubsection{LED energy width}
\begin{figure}
			\includegraphics[scale=0.34]{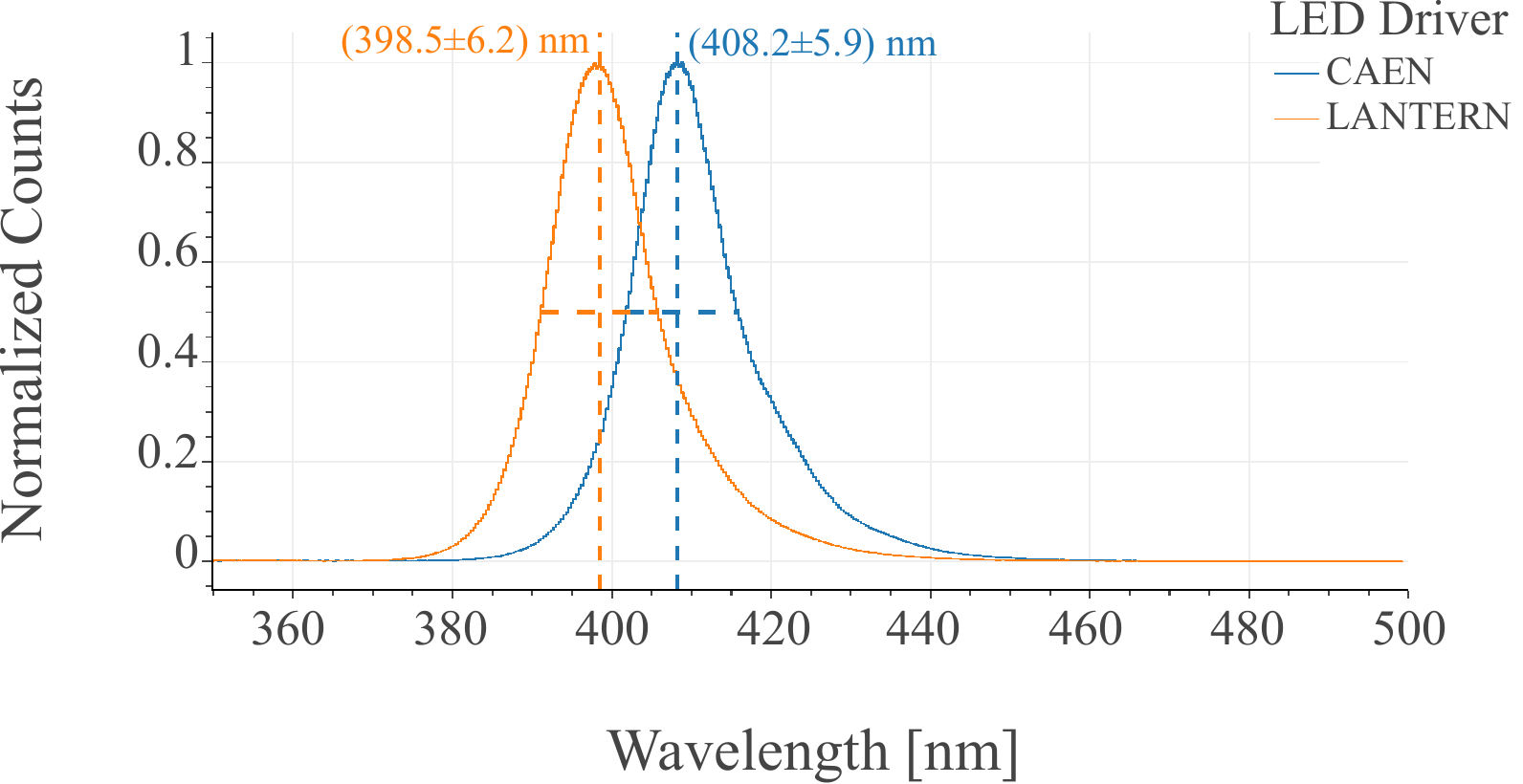}
			\caption{Optical spectrum of the photons emitted by different LED sources. In orange the LLS-UV400 LED used in the LANTERN electronics, while in blue the SP5601 CAEN commercial LED driver are shown. Both spectra have been measured using a PyLoN:100BR\_eXelon CCD~\cite{pylon} and renormalized for the quantum efficiency of the device.}
			\label{fig:LED_spectrum}
	     \end{figure}

		As shown in Fig.~\ref{fig:LED_spectrum} an LED is not completely monochromatic but usually presents a $\approx 2\%$ width in the wavelength distribution. This means that Eq.~\ref{eq:abscal_func} needs to be corrected to include this effect and thus becomes:
		
\begin{multline}
\sigma^2_{LED} = \sigma_0^2+\sigma^2_\gamma+\sigma^2_{\langle\epsilon_\gamma\rangle}r^2\langle N_\gamma\rangle^2 =  \sigma_0^2+\sigma^2_\gamma+\sigma^2_{\epsilon_\gamma}r^2\langle N_\gamma\rangle =  \\
=\sigma_0^2+\sigma^2_\gamma + r^2 \langle\epsilon_\gamma\rangle^2\langle N_\gamma\rangle \frac{\sigma^2_{\epsilon_\gamma}}{\langle\epsilon_\gamma\rangle^2}= \sigma_0^2+\sigma^2_\gamma \left(1+\frac{\sigma^2_{\epsilon_\gamma}}{\langle\epsilon_\gamma\rangle^2}\right)
\end{multline}
where $\sigma^2_{\langle\epsilon_\gamma\rangle} = \frac{\sigma^2_{\epsilon_\gamma}}{\langle N_\gamma\rangle}$ is the variance of the mean photon energy. The above equation shows that the correction induced by the fact that the LED is not perfectly monochromatic scales quadratically with the relative width of the photon energy distribution and is thus of the order of $10^{-4}$. In practice, this correction is negligible since the typical statistical error on the calibration parameters are at the order of few percent, and thus justifies the use of LEDs instead of laser diodes.
 
	\subsection{Nonlinearity evaluation}\label{sec:nonlin}
	As shown in Fig.~\ref{fig:Simu_pulse}, shining the photons via a series of equal bursts allows for a linear control of $\langle N_\gamma \rangle$ and thus a linear control of $\mu_\gamma$. Having this degree of control on the deposited energy, while not required by the calibration itself, can be used to correct possible detector nonlinearities. 
	
\begin{figure}
			\includegraphics[scale=0.35]{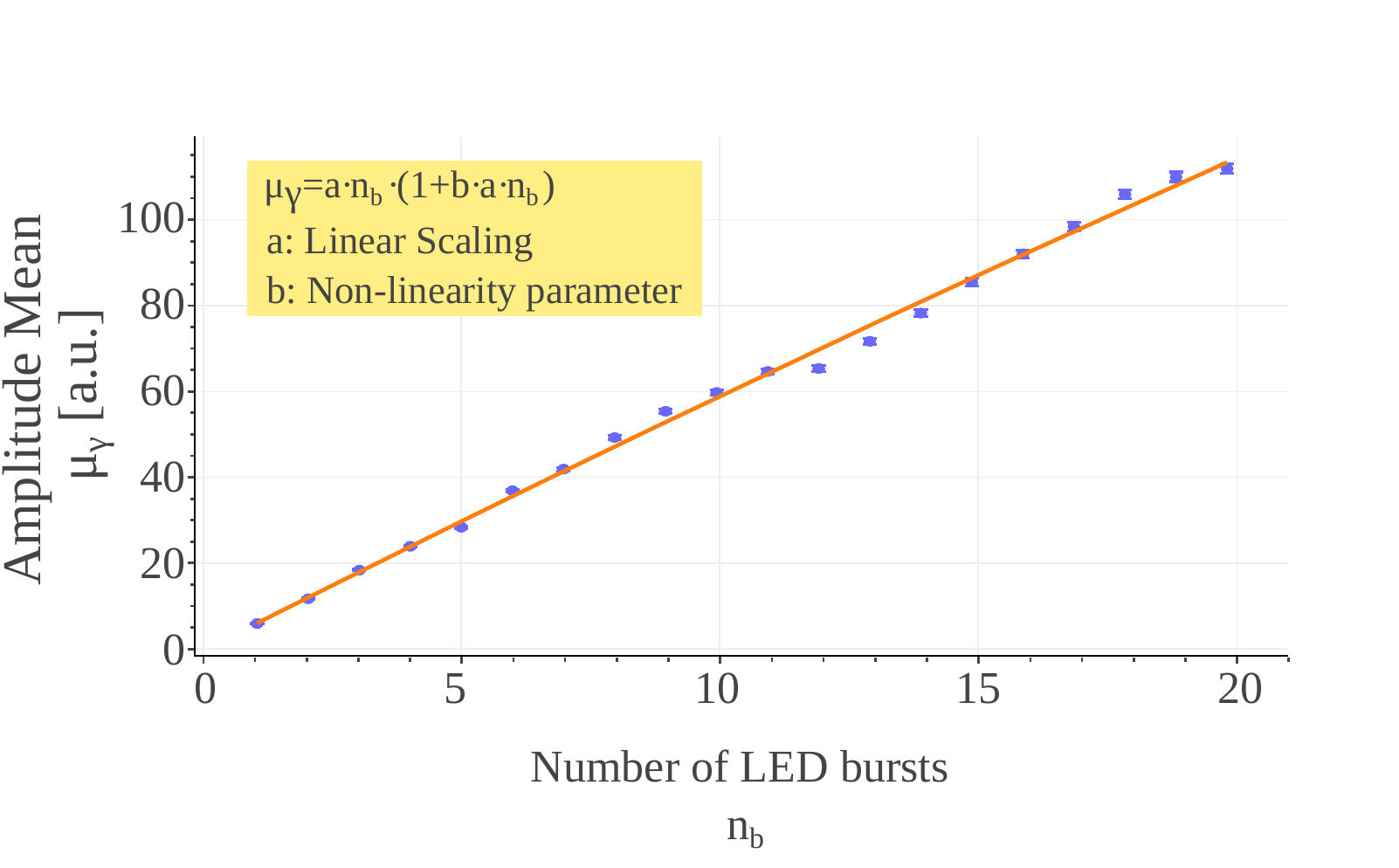}
			\caption{Example of the estimation of the nonlinearity of the detector using the optical calibration setup. The mean of the energy depositions of Fig.~\ref{fig:calib_example} is plotted against the number of bursts used to produce the distribution. The data points are then fitted with Eq.~\ref{eq:non_linear} to extract the energy nonlinearity correction parameters.}
			\label{fig:linearization_example}
		 \end{figure}

	In fact, if $n_b\propto\langle N_\gamma\rangle$ is the number of light bursts it is possible to correct for higher order dependencies of the mean amplitude of the depositions with respect to the delivered energy. In absence of a response model of the detector it is possible to study the next leading order, quadratic, in the dependence of the amplitude with respect to the deposited energy, meaning that the response becomes:
	\begin{equation} \label{eq:non_linear}
	\mu_\gamma=a\cdot n_b \left(1+b\cdot a \cdot n_b\right)
	\end{equation}
where $a$ is the linear scaling parameter and $b$ encodes the relative quadratic dependence on the delivered energy. When fitting the above equation to the $\mu_\gamma(n_b)$ scaling shown in Fig.~\ref{fig:linearization_example}, it is possible to extract the $a$ and $b$ parameters to then correct the nonlinearity. In fact, once these parameters are known the linearly scaling amplitude variable can be evaluated as:
\begin{equation}
A_L = \frac{-1 + \sqrt{1+4b\cdot A}}{2b}
\end{equation}
where $A$ is the amplitude of the raw detector pulse and $A_L$ is the corrected amplitude that scales linearly with the deposited energy. It can be shown that for LED induced pulses the mean amplitude $\langle A_L\rangle$ scales linearly with $n_b$.

Since the procedure described in Sec.~\ref{sec:calib} assumes the linear scaling between pulse amplitude and energy, it is advisable to perform the linearization procedure before performing the optical calibration.

\section{Electronics Design}
As already mentioned, the typical integration time of a cryogenic calorimeter is equal or greater than few hundred microseconds, meaning that the LED bursts must be delivered in at most 10~$\mu$s in order to not produce any pulse shape deformation. To have a wide dynamic range of characterization and also high granularity in the energy deposition a target number of the maximum cycles to be delivered is 100. This means that a single cycle must be around 100~ns long, i.e. the LED driving circuit must be designed to have switching frequencies of a few MHz. The targeted frequency for the developed electronics is chosen to be 5~MHz, but can be tuned to the specific experimental needs.

The LANTERN electronics is designed to fulfill the above requirement and the following two characteristics that concern the ease of up-scaling the project:
\begin{itemize}
\item modular and extendable electronics in order to calibrate large arrays of cryogenic calorimeters;
\item vacuum compatible electronics for operation inside the cryostat vessels.
\end{itemize}

\begin{figure}
		\includegraphics[scale=0.35]{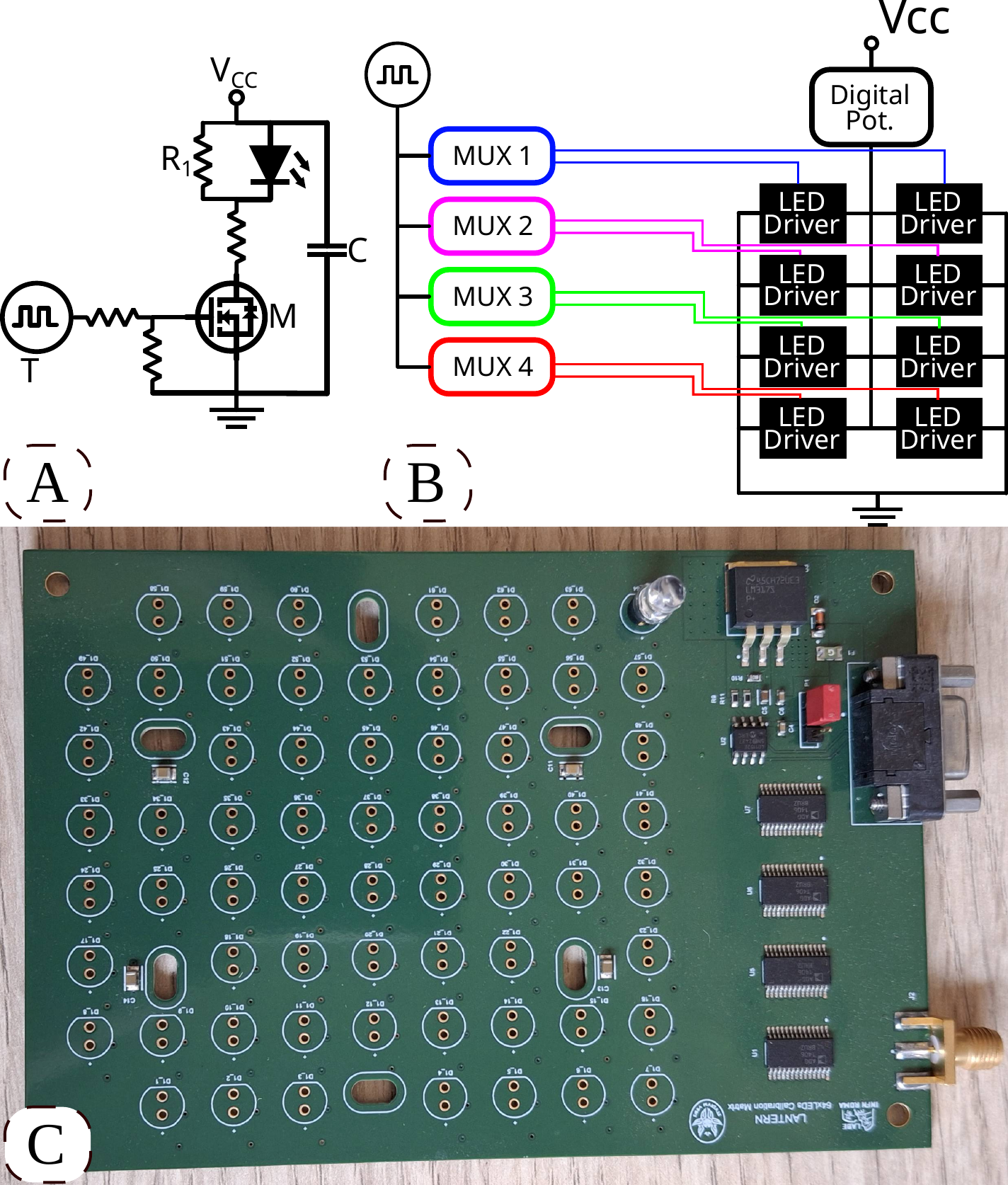}
		\caption{Development of the LANTERN optical calibration electronics. Panel A: design of the single LED driver developed around the use of the BSS123 MOSFET (M) to bias the LED with an externally generated triggering signal T. The R$_1$ resistor is used to speed up the discharge of the LED and the capacity C is used to protect the circuit from possible high frequency fluctuations of the bias voltage V$_\text{CC}$. Panel B: conceptual design of the multiplexing scheme used to up-scale the number of optical channels. Four multiplexers are employed to redirect the triggering signal to one of the available LED drivers, which are all connected in parallel to the bias voltage. A digital potentiometer is used to accurately regulate the voltage reaching the LED driver, making the luminosity easily adjustable. Panel C: Picture of the final printed circuit board of LANTERN with 64 channels. Only one LED is mounted to perform the initial validation of the electronics.}
		\label{fig:design}
	 \end{figure}

The design of the single LED driving unit is shown in panel A of Fig.~\ref{fig:design}. The LED driving circuit is sent into conduction via the use of the BSS123 MOSFET, chosen for the high speed switching and maximum current, coupled to an externally generated triggering signal. The other main feature of the driving circuit is the resistor $R_1$ placed in parallel with the LED, which is used to increase the discharge speed of the LED in order to reduce the switching time of the driver. Low values of the resistor $R_1$ can increase the switching speed of the driver but reduce the intensity of the LED while increasing the resistance will decrease the switching frequency of the circuit. For this reason a value around 200~$\Omega$ is found to be a good bargain for most commercial LEDs but in an optimal configuration this value must be tuned considering the capacity of the LED. Tests of the single LED driver circuit were performed in the initial stages of the project and are described in \cite{lantern1}.

To up-scale the number of optical channels to 64 LED drivers, a multiplexing scheme has been devised based on the use of the ADG1406BRUZ multiplexer. As visible from panel B of Fig.~\ref{fig:design}, the multiplexer redirects the triggering signal to the chosen LED driver that needs to be biased, while the biasing power V$_\text{cc}$ is always distributed to all the channels. An additional feature of the multiplexed electronics is the integration of a digital potentiometer used to regulated the biasing voltage of each LED in order to tune their luminosity to the required value. The final printed circuit board (PCB), visible in panel C of Fig.~\ref{fig:design}, is produced without any on-chip microcontroller, in this way the system can be easily integrated in the DAQ systems of the experiments that employ LANTERN. For the initial tests of the system a communicating module based on the use of an Arduino Nano was developed to act as interface with the laboratory computer.

\section{Electronics validation in the laboratory}
\begin{figure}
	\includegraphics[scale=0.4]{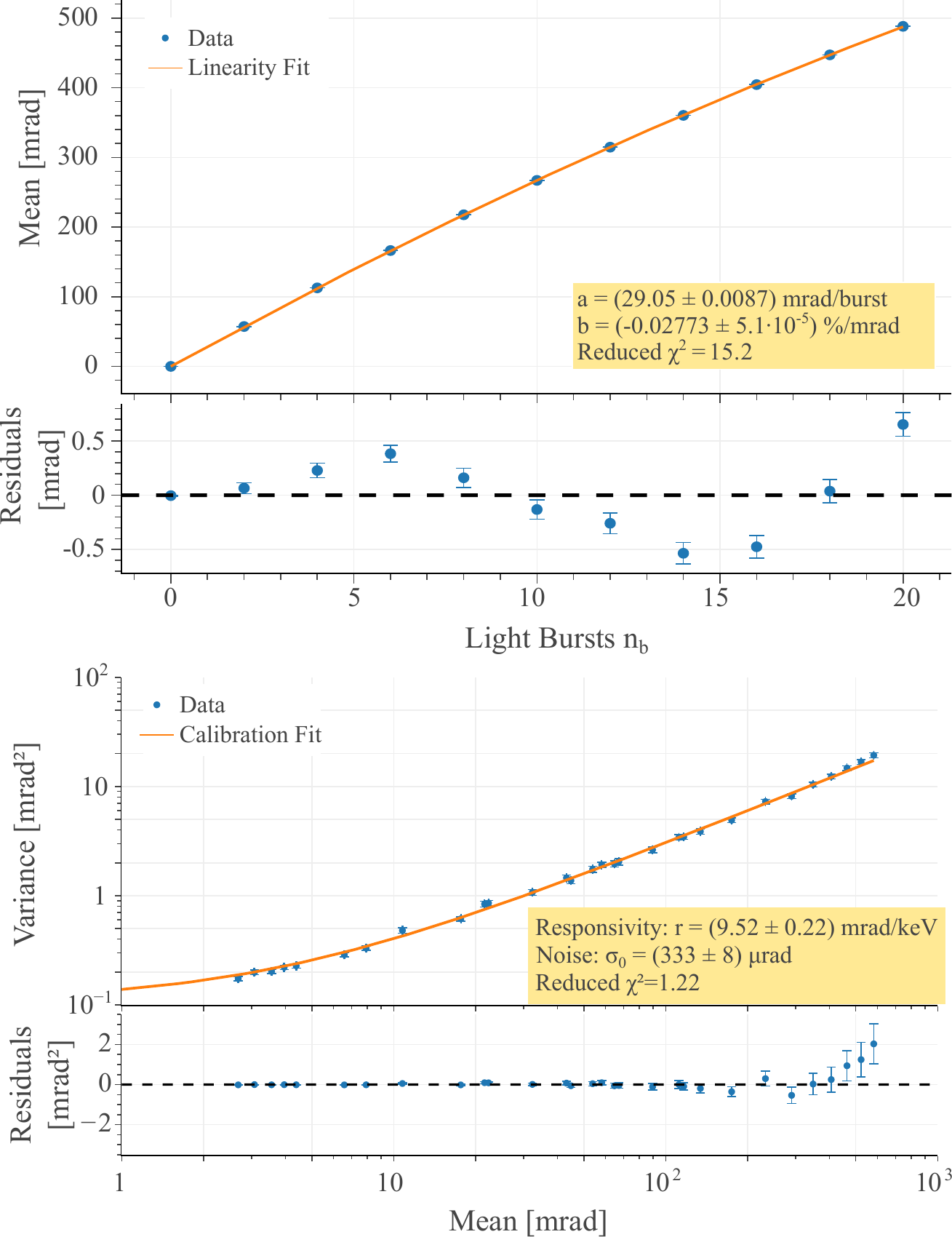}
	\caption{Evaluation of the detector response of the BULLKID-DM \cite{bullkid} detector used in the measurement of the spectrum in Fig.~\ref{fig:room_temp_cal}. In the top panel the nonlinearity fit is performed as described in Sec.~\ref{sec:nonlin}. The high $\chi^2$ is the do to the lack of a response model of the detector, but the general trend of the linearity is well described by the fitted second degree polynomial. The bottom panel shows the optical calibration performed after the linearization of the detector response. The plot is presented in an unusual log-log scale to emphasize the wide energy range of the calibration that goes from around 300~eV to 60~keV. The bottom plot has more data points than the top one since less measurements are required to reach the desired level of precision in the linearization of the detector response.}
		\label{fig:room_temp_optical}
\end{figure}

After the success of the first tests performed with the developed electronics, the LANTERN system was used to calibrate one of the cryogenic detector arrays employed by the BULLKID-DM experiment \cite{bullkid}. The sensors used to obtain the results in the rest of this work are phonon-mediated kinetic inductance detectors (KIDs), which consist of a superconductive LC resonator with resonant frequency of around 1~GHz. The observable signal from a KID is a shift in the resonance frequency of the sensor which is typically recorded by monitoring a phase change in the transmission at a fixed frequency (see \cite{bullkid} for more details) and the signals are thus measured in milliradiants. The calibration was performed as described in sections \ref{sec:calib} and \ref{sec:nonlin} and is shown in Fig.~\ref{fig:room_temp_optical}. The top panel of the figure presents the linearization of the detector response up to 60~keV for which a few points are required to reach the desired precision in the procedure since it is limited by the lack of a more accurate model of the detector response. In the bottom panel the optical calibration fit is presented, it is worth noticing that the calibration is performed starting from 300~eV up to 60~keV thus characterizing of the full dynamic range of the sensor.

\begin{figure}
	\includegraphics[scale=0.38]{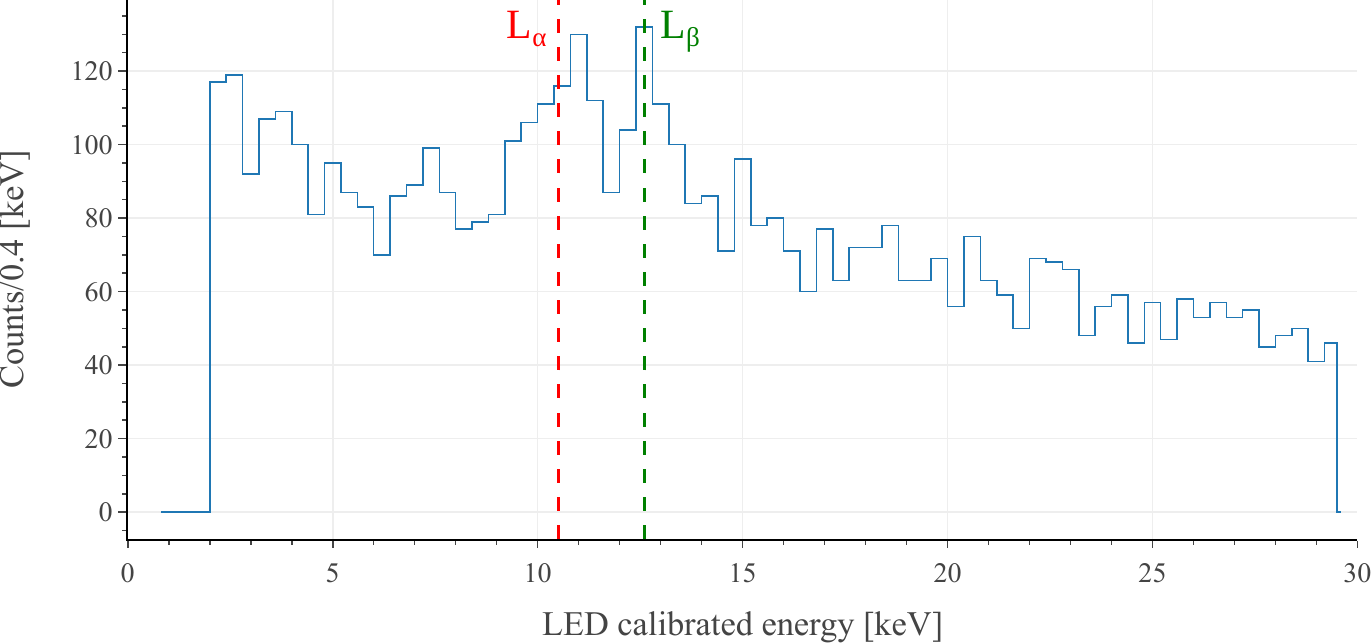}
	\caption{Cross-evaluation of the LED calibration with the peaks induced by the lead casing surrounding the detector. The plot shows the optically calibrated energy spectrum measured by the BULLKID detector in the [2,30]~keV range. In the spectrum the L-shell $\alpha$ (10.5~keV) and $\beta$ (12.6~keV) X-ray lines are present. The average reconstructed energy of said lines are respectively $(10.7\pm 0.1)$~keV and $(12.8\pm 0.1)$~keV.}
		\label{fig:room_temp_cal}
\end{figure}

In the resulting energy spectrum of the measurement, which is object of an upcoming publication by the collaboration, two X-ray peaks produced from de-excitation of the atoms in the lead case surrounding the detector can be seen as visible from  Fig.~\ref{fig:room_temp_cal}. The discrepancy between the optical calibration and the nominal peak positions is estimated to be at the level of $2\%$.

The good agreement in the energy reconstruction between the measured peaks and the optical calibration validates the use of the procedure for the characterization of cryogenic calorimeters. Moreover, it proves the potential and accuracy of the LANTERN electronics when used to calibrate and characterize the response of cryogenic detectors.

\section{Operation in vacuum}
After the confirmation of the correct behavior of the electronics when deployed outside the cryostat vacuum shield, the LANTERN PCB in Fig.~\ref{fig:design} was installed inside the cryostat at the 300~K stage, i.e. in vacuum but at room temperature, coupled to a heater in order to compensate for the temperature decrease due to radiative losses. The heater proved to be necessary since the PCB would otherwise reach -20° due radiation losses, and showed problematic behavior; a stable operating temperature of 20°C was chosen and reached with the heater.

\begin{figure}
	\includegraphics[scale=0.52	]{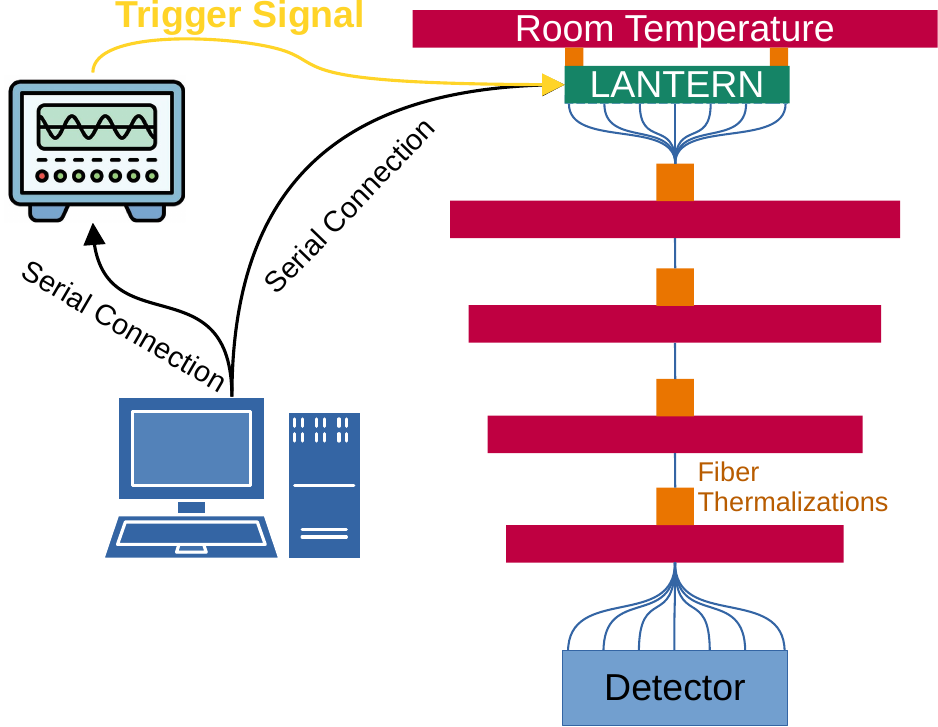}
	\caption{Conceptual design of the final deployment of the LANTERN calibration system inside the vacuum of the cryostat. In red the cryostat plates are shown, while in green the LANTERN PCB is presented deployed at the room temperature stage inside the vacuum vessel. The optical fibers that bring the photons from the electronics to the detector are thermalized at each cryogenic stage by a custom designed copper piece. The control of the electronics is done in the laboratory environment via the acquisition computer which communicates with a signal generator, to produce the trigger, and with the LANTERN board, for the slow control dedicated to the selection of the LED driver.}
		\label{fig:vacuum_setup}
	\end{figure}

The full system design, shown in Fig.~\ref{fig:vacuum_setup}, has the PCB deployed inside the cryostat and thermalized at room temperature. The board is then connected to a bundle of optical fibers which is guided through the cryostat plates via the use of a series of dedicated thermalizations and arrives to the detector where each fiber is coupled to one element of the cryogenic calorimeter array of the BULLKID-DM \cite{bullkid} experiment. 

The choice of deploying the electronics at the room temperature stage of the cryostat was driven by two main considerations. The first one is that at colder stages the LED will change the wavelength of the emitted photons due to the temperature dependence, and thus requires a dedicated study of the produced optical spectrum. The second consideration is that circuit boards are known to be quite dirty in terms of radiopurity, thus placing the electronics far from the detector while also leaving the possibility to place a radiation shield in the between the two is a desirable feature for the low background experiments for which this setup is designed for.

From outside of the cryostat the triggering signal, a 5~MHz square wave, is produced via a dedicated signal generator piloted with the acquisition computer. The same computer is also coupled to the PCB for the slow control used to choose the desired optical channel. This setup concept is currently being deployed in Università "La Sapienza" in Rome (Italy) in the R\&D facility for the BULLKID-DM \cite{bullkid} detectors.

\subsection{Cross-calibration with commercial system}
To fully validate the proper operation of the system deployed inside the cryostat, one of the cryogenic calorimeters produced by the CALDER \cite{calder} project was used. The wide area of the calorimeter (2~in $\times$ 2~in) allowed to place two optical fibers, one coupled to one of LANTERN's LEDs and the other to the SP5601 400~nm CAEN LED driver operated in the laboratory environment. This commercial driver has been used as the baseline version of the optical calibration setup. 

\begin{figure}
	\includegraphics[scale=0.44]{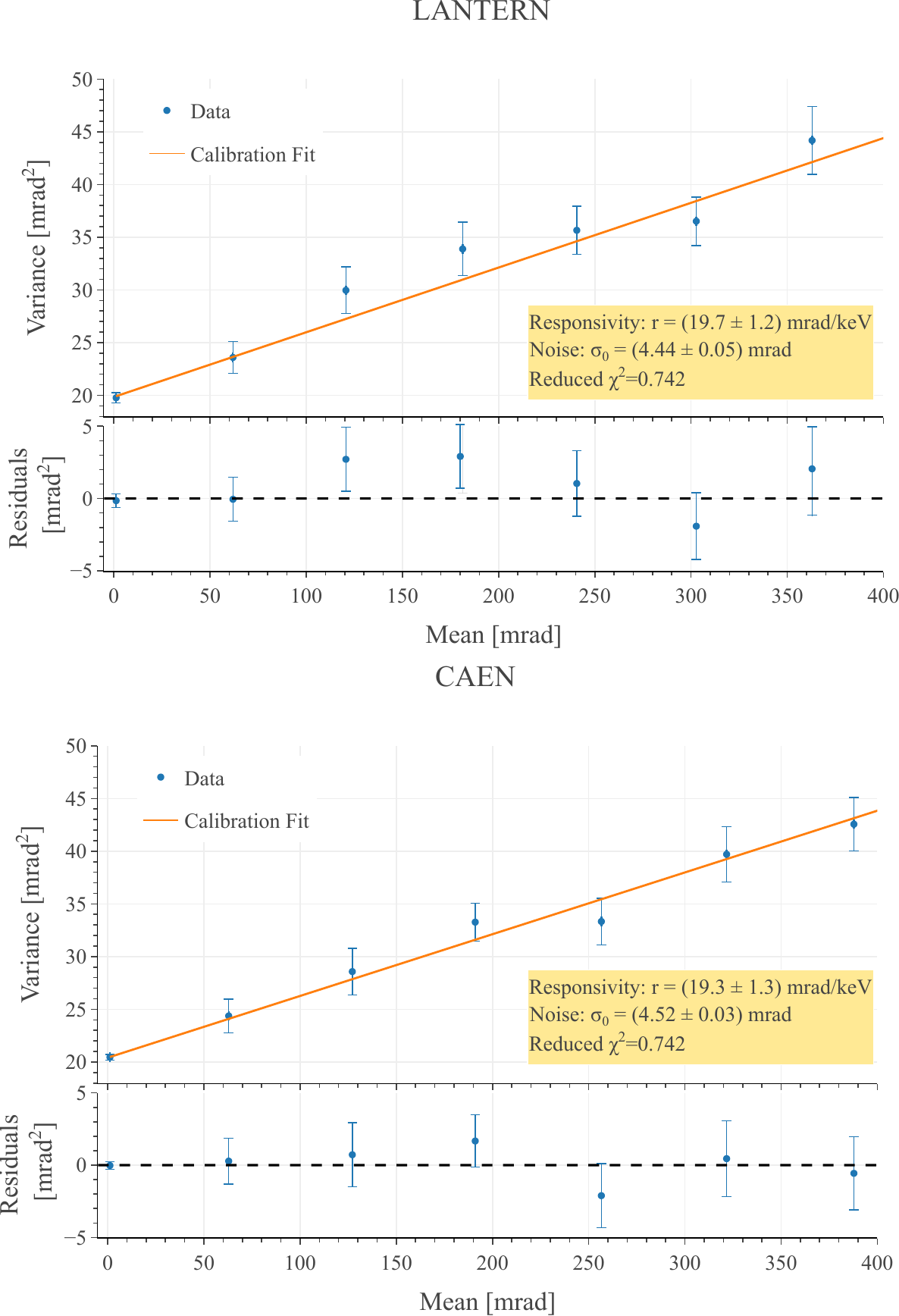}
	\caption{Validation of the correct behavior of the LANTERN electronics when deployed in vacuum. The two panels show the results of two optical calibrations performed respectively with the LANTERN electronics deployed inside the vacuum of the cryostat vessel and the SP5601 CAEN commercial LED driver placed in the laboratory environment. The two systems are used to calibrate one of the cryogenic detectors developed by the CALDER project \cite{calder}. The two calibrations yield compatible results fully validating the LANTERN electronics.}
	\label{fig:vacuum_cal}
	\end{figure}

As visible from Fig.~\ref{fig:vacuum_cal} the two different calibrations with the two systems converged at compatible calibration parameters, both in terms of responsivity and intrinsic noise level. This validates the operation of the electronics in the vacuum of the cryostat removing the need of vacuum optical feedthroughs to bring the light to the detector units, thus greatly simplifying the setup.

\section{Discussion and Perspectives}
The LANTERN calibration system has been developed in phases: the first phase of the development of the single LED driving circuit has been validated in \cite{lantern1} and in this work the validation of the up-scaled electronics is shown. 

LANTERN showed the potential to achieve an accurate calibration both in the low energy regime, few hundred electronvolts, up to several tens of keV. With this wide energy range, it was possible to compare the optical calibration with the peaks produced by the lead casing surrounding the detector. This comparison showed a less than 2\% discrepancy in the energy reconstruction. 

Moreover, it was demonstrated that the electronics and calibration setup is easily up-scaled to a large amount of optical channels with the current target of 64 due to the synergy with the BULLKID-DM \cite{bullkid} detectors. To up-scale this setup it was imperative to design an electronics capable of operating inside the vacuum vessel of the cryostat. The correct behavior of the setup in vacuum was validated with a cross-calibration done using a simpler commercial setup deployed in the laboratory environment. In the operation in vacuum a way to heat the circuit board to the desired value of 20°C has proven to be crucial for the correct operation of the device and to achieve the desired quality of the calibration.

At the moment, LANTERN is being considered as the calibration and monitoring electronics for the BULLKID \cite{bullkid}, CRAB \cite{crab} and NUCLEUS \cite{NUCLEUS} experiments requiring minimal variations, mostly in terms of LED wavelength, between the various versions to have a better optical matching with crystal substrate used by the experiment.

\section{Acknowledgments}
The realization of this project was only possible with the precious help of A. Girardi and L. Recchia in designing the electronics, of A. Cruciani and M. Vignati for overall guidance, R. Cerulli for the availability during the prototype validation, A. Polimeni and S. Cianci for the optical validation of the components and D. Pasciuto for the design of the thermalization strategy. This project was funded by the ‘‘Progetti per Avvio alla ricerca’’ of the Sapienza University and also supported by the European Union (ERC, DANAE, 101087663). Views and opinions expressed are however those of the author(s) only and do not necessarily reflect those of the European Union or the European Research Council. Neither the European Union nor the granting authority can be held responsible for them.

\end{document}